\documentclass[12pt]{iopart}

\usepackage{graphicx}
\usepackage[T1]{fontenc}
\usepackage{epsfig} 
\usepackage{ulem}
\usepackage{amssymb}
\usepackage[mathscr]{eucal}
\usepackage{hyperref}
\usepackage{color}
\usepackage{bbold}

\newcommand{\eqref}[1]{(\ref{#1})}

\begin{document}

\title{From quantum to classical modelling of radiation reaction: a focus on the radiation spectrum}
\author{F. Niel$^{1,\dagger}$, C. Riconda$^1$, F. Amiranoff$^1$, M. Lobet$^2$, J.~Derouillat$^2$, F. P\'erez$^3$, T. Vinci$^1$ and M. Grech$^{3,\star}$}
\address{$^1$LULI, UPMC Universit\'e Paris 06: Sorbonne Universit\'es, CNRS, \'Ecole Polytechnique, CEA, Universit\'e Paris-Saclay, F-75252 Paris cedex 05, France}
\address{$^2$Maison de la Simulation, CEA, CNRS, Universit\'e Paris-Sud, Universit\'e Paris-Saclay, F-91191 Gif-sur-Yvette, France}
\address{$^3$LULI, CNRS, \'Ecole Polytechnique, CEA,  Universit\'e Paris-Saclay, UPMC Universit\'e Paris 06: Sorbonne Universit\'es, F-91128 Palaiseau cedex, France}
\ead{$^{\dagger}$fabien.niel@polytechnique.edu, $^{\star}$mickael.grech@polytechnique.edu}

\begin{abstract}
Soon available multi petawatt ultra-high-intensity (UHI) lasers will allow us to probe high-amplitude electromagnetic fields interacting with
either ultra-relativistic electron beams or hot plasmas in the so-called moderately quantum regime. 
The correct modelling of the back-reaction of high-energy photon emission on the radiating electron dynamics, 
a.k.a. radiation reaction, in this regime is a key point for UHI physics. 
This will lead to both validation of theoretical predictions on the photon spectrum emitted during the laser-particle interaction and 
to the generation of high energy photon sources. 
In this paper we analyse in detail such emission using recently developed models to account for radiation reaction.
We show how the predictions on the spectrum can be linked to a reduced description of the electron distribution function in terms of the first energy moments. 
The temporal evolution of the spectrum is discussed, as well as the parameters for which quantum effects induce hardening of the spectrum. 
\end{abstract}

\maketitle

\section{Introduction}\label{sec:intro}
Radiation reaction (RR) is the influence of the electromagnetic field emitted by a charged particle on its own dynamics. 
It is one of the oldest problems in electrodynamics~\cite{Landau_CED,Spohn,Di_Piazza_RMP}, it has recently been at the center of various studies 
motivated either by relativistic astrophysics studies~\cite{jaroschek2009,cerutti2014,cerutti2016}, or by the advent of multi-petawatt laser systems~\cite{tamburini2010,sokolov2010,duclous2011,nerush2011,ridgers2012,capdessus2013,blackburn2014,gonoskov2015,lobet2015,lobet2017,grismayer2017}.
Among the latter, the Apollon~\cite{Apollon}, Gemini~\cite{cole2017,poder2017} and ELI~\cite{ELI} facilities provide us with test beds for our models 
of high-energy photon emission and its back-reaction on the particle dynamics.

In order to provide a link between theoretical/numerical modelling and experiments, many authors have studied the signatures of RR on either
the electron or photon distributions and proposed accordingly possible experimental setups in order to observe such signatures. 
Among the signatures in the electron distribution function, one can mention the so-called (quantum) stochastic heating first predicted by Neitz and Di Piazza~\cite{DiPiazza_Neitz}
and further investigated in Refs.~\cite{vranic2016,FNiel,ridgers2017}
or the so-called quenching of radiation losses first introduced in Ref.~\cite{harvey2017} and more recently revisited as an asymmetry (or skewness) of the electron distribution 
function in Ref.~\cite{FNiel}. 

Signatures in the emitted photon spectrum have also been studied. 
RR was identified with the multiple photon recoils experienced by electrons emitting consecutively and incoherently high-energy photons in Ref.~\cite{dipiazza2010},
a study in which the authors also highlighted the impact of  RR on the multi-photon Compton spectra.
More recently, Neitz and Di Piazza also used a kinetic approach to study the interplay of RR and quantum electrodynamics (QED) effects on both the electron and emitted photon spectra,
suggesting that such effects could be observed at already available laser intensities~\cite{NeitzKinetic}.
Such findings were further supported by Blackburn {\it et al.}, who used a Monte-Carlo based Particle-In-Cell (PIC) approach to study high-energy photon emission in the head-on collision
of a electron-beam with an ultra-intense laser pulse in a regime where QED are important~\cite{Blackburn}.

In the present paper, we investigate RR effects on the spectrum of high-energy photons focusing on the comparison of the three different models discussed in Ref.~\cite{FNiel}.
The paper is structured  as follows. 
First, Sec.~\ref{sec:rad_spectrum} lays the basis for this study. 
We summarize previous results on the radiation spectrum considering a single radiating electron in the so-called local constant field approximation. 
We then introduce the three complementary approaches for treating RR in the moderately quantum regime which have been recently presented in Ref.~\cite{FNiel} 
and discuss how the emitted high-energy photon properties can be extracted using these approaches. 
Then, in Sec.~\ref{sec:results}, we present the results of PIC simulations performed with the open-source code {\sc Smilei}~\cite{smilei}.
These simulations shed a new light on how the three models for RR mentioned above perform in predicting the spectral properties of the emitted high-energy photons.
Finally, Sec.~\ref{sec:conclusion} summarizes our findings.

\section{High-energy photon emission and its back reaction}\label{sec:rad_spectrum}

Throughout this work, we place ourselves within the local constant field approximation (LCFA) which relies on the possibility
of neglecting the space and time variation of the external field over the region of formation of a high-energy photon 
emitted by an ultra-relativistic electron. The LCFA validity has been investigated in recent works~\cite{dipiazza_arxiv_2017}, 
but its discussion remains beyond the scope of this work.

In what follows, we first recall the properties of high-energy photon emission under the LCFA, 
focusing on the single electron emission rate and spectral properties of the emitted photons (Sec.~\ref{sec:rad_spectrum:A}).
We then summarize previous findings on the modelling of the back-reaction of high-energy photon emission on the radiating electron dynamics, a.k.a. radiation reaction, 
in the so-called moderately quantum regime (Sec.~\ref{sec:rad_spectrum:B}). 
Finally, in Sec.~\ref{sec:rad_spectrum:C}, we take a closer look at the high-energy photon emission by an ensemble of ultra-relativistic electrons. 

\subsection{Photon emission rate and energy spectrum}\label{sec:rad_spectrum:A}

Let us consider an electron (with charge $-e$ and mass $m$) radiating in an external electromagnetic field with relativistic
strength:
\begin{eqnarray}
a_0 = \frac{e \vert A^{\mu}\vert}{m c^2} \gg 1,
\end{eqnarray}
where $c$ is the speed of light [SI units will be used throughout this work unless specified otherwise],
and $A_{\mu}$ is the four potential associated to the electromagnetic field tensor $F^{\mu\nu}=\partial^{\mu}A^{\nu}-\partial^{\nu}A^{\mu}$. 
We will further consider undercritical (otherwise arbitrary) electromagnetic fields,
that is both field invariants are small with respect to the square of the Schwinger field $E_s = m^2 c^3/(e\hbar) \simeq 1.3\times 10^{18}~{\rm V/m}$.
The Lorentz invariant rate of photon emission by the electron has been first derived in the LCFA 
by Nikishov and Ritus~\cite{Nikishov}. It reads:
\begin{eqnarray}\label{eq:emission-rate}
\frac{d^2N_{\gamma}}{d\tau d\chi_{\gamma}} = \frac{2}{3}\frac{\alpha^2}{\tau_e}\,\frac{G(\chi,\chi_{\gamma})}{\chi_{\gamma}}\,,
\end{eqnarray}
where $\tau$ is the electron proper time, $\alpha = e^2/(4\pi\epsilon_0 \hbar c)$ is the fine structure constant 
($\epsilon_0$ and $\hbar$ being the permittivity of vacuum and reduced Planck constant, respectively), 
$\tau_e = r_e/c$ is the time for light to cross the classical radius of the electron $r_e = e^2/(4\pi\epsilon_0 m c^2)$, 
and $G(\chi,\chi_{\gamma})$ is the so-called quantum emissivity:
\begin{eqnarray}
G(\chi,\chi_{\gamma}) = \frac{\sqrt{3}}{2\pi}\frac{\chi_{\gamma}}{\chi}\left[ \int_{\nu}^{+\infty}\!\!\!\! K_{5/3}(y)\, dy + \frac{3}{2} \chi_{\gamma} \nu \, K_{2/3}(\nu) \right]
\end{eqnarray}
with $\nu = 2\chi_{\gamma}/[3\chi(\chi-\chi_{\gamma})]$. 
Both the photon emission rate and the quantum emissivity depend only on the electron and photon parameters at the moment of emission:
\begin{eqnarray}
\chi = \left\vert \frac{F^{\mu\nu}}{E_s}\frac{p_{\nu}}{m c} \right\vert \quad {\rm and} \quad 
\chi_{\gamma} = \left\vert \frac{F^{\mu\nu}}{E_s}\frac{\hbar k_{\nu}}{m c} \right\vert
\end{eqnarray}
with $p^{\mu}$ and $\hbar k^{\mu}$ the electron and photon four-momentum, respectively.

Let us now consider a reference frame, henceforth referred to as the laboratory-frame, in which the electron is ultra-relativistic (that is the electron Lorentz factor is $\gamma \gg 1$). 
The photon quantum parameter at the time of emission can be directly linked to the electron quantum parameter at this time 
as $\chi_{\gamma}=\xi\,\chi$, with $\xi = \gamma_{\gamma}/\gamma$ the ratio of the photon normalized energy $\gamma_{\gamma} = \hbar\omega/(m c^2)$ ($\omega$ denoting the photon angular frequency) by the electron Lorentz factor. 
The instantaneous power spectrum of the emitted high-energy photons (in this frame, and for a single radiating electron) is obtained from Eq.~\eqref{eq:emission-rate} and reads:
\begin{eqnarray}\label{eq:spectrum}
\frac{dP_{\rm inst}}{d\gamma_{\gamma}} =  \frac{\sqrt{3}}{2\pi} P_{\alpha} \, \frac{\xi}{\gamma} \left[ \int_{\nu}^{+\infty}\!\!\!\! K_{5/3}(y)\, dy + \frac{\xi^2}{1-\xi} \, K_{2/3}(\nu) \right]\,,
\end{eqnarray}
with $P_{\alpha} = 2 \alpha^2 mc^2/(3\tau_e)$. 
Let us note that, in the classical limit $\chi \ll 1$, Eq.~\eqref{eq:spectrum} reduces to the standard synchrotron energy spectrum (see, e.g., Ref.~\cite{jackson}), 
and integration of Eq.~\eqref{eq:spectrum} over all photon energies leads the {\it quantum corrected} instantaneous power radiated away by the electron:
\begin{eqnarray}
P_{\rm rad} = P_{\rm cl}\, g(\chi)\,,
\end{eqnarray}
with $P_{\rm cl} = P_{\alpha} \chi^2$ the classical Larmor power and 
\begin{eqnarray}
\label{eq:g:quantum-correction} g(\chi) =  \frac{9\sqrt{3}}{8\pi}\!\!
 \int_0^{+\infty}\!\!\!\!\! d\nu\,\left[ \frac{2\nu^2\,{\rm K}_{5/3}(\nu)}{(2+3\nu\chi)^2}  
 +  \frac{4\nu\, (3\nu\chi)^2}{(2+3\nu\chi)^4}\, {\rm K}_{2/3}(\nu)\right]\,
\end{eqnarray}
the quantum correction introduced in various works~~\cite{ridgers2014,Erber} (see also Ref.\cite{FNiel} for more details). 
Let us finally stress that considering ultra-relativistic electrons further allows us to assume that all photons are emitted in the direction ${\bf \Omega}$ 
of the electron velocity at the moment of emission.  

\subsection{Radiation reaction}\label{sec:rad_spectrum:B}

In the moderately quantum regime~\cite{Di_Piazza_RMP} ($a_0\gg 1$ and $\chi \lesssim 1$), electron-positron pair production can be neglected and
the problem of radiation reaction acting on a population of ultra-relativistic electron can be casted in the form of a Master 
equation for the electron and photon distribution functions $f_e(t,{\bf x},\gamma,{\bf \Omega})$ and 
$f_{\gamma}(t,{\bf x},\gamma,{\bf\Omega})$, respectively~\cite{sokolov2010,FNiel,ridgers2017,Elkina,Neitz}:
\begin{eqnarray}
\nonumber\!\!\!\!\frac{d}{dt} f_e &=& \,\int_0^{+\infty}\!\!\!\!d\gamma_{\gamma}\,w_{\chi} (\gamma+\gamma_{\gamma},\gamma_{\gamma})  \, f_e(t,{\bf x,}\gamma + \gamma_{\gamma},{\bf\Omega}) \\
\label{eq:Master1} &-& f_e(t,{\bf x,}\gamma,{\bf\Omega})\,\int_0^{+\infty}\!\!\!\!d\gamma_{\gamma} w_{\chi} (\gamma,\gamma_{\gamma})\,,\\
\label{eq:Master2}\!\!\!\! \frac{d}{dt} f_{\gamma} &=& \int_1^{+\infty}\!\!\!\!d\gamma\, w_{\chi}(\gamma,\gamma_{\gamma})\,f_e(t,{\bf x,}\gamma,{\bf\Omega}),
\end{eqnarray}
where $w_{\chi}(\gamma,\gamma_{\gamma})$ is the rate of emission of photons with energy $m c^2\gamma_{\gamma}$ by an electron 
with a given quantum parameter $\chi$ and energy $m c^2 \gamma$:
\begin{eqnarray}
w_{\chi}(\gamma,\gamma_{\gamma}) = \frac{2}{3}\frac{\alpha^2}{\tau_e}\frac{G(\chi,\chi_{\gamma})}{\gamma \gamma_{\gamma}}\,,
\end{eqnarray}
and where it has been assumed that radiation emission (and its back-reaction) is dominated by
the contribution of ultra-relativistic electrons (for which ${\bf p}\simeq mc \gamma {\bf\Omega}$),
and that such ultra-relativistic electrons emit radiation in the direction $\bf\Omega$ of their velocity.
Note also that the time derivative $d/dt$ in the rhs side of Eqs.~\eqref{eq:Master1} and~\eqref{eq:Master2}
are total time derivatives and have to be handled carefully.

This Master equation describes the process of high-energy photon emission as a discontinuous jump process. 
Equation~\eqref{eq:Master1} is a linear Boltzmann equation whose rhs acts as a {\it collision operator},
while the rhs of Eq.~\eqref{eq:Master2} is a source term for the photon population.
A discrete formulation of this Master equation can be rigorously derived in the form of the Monte-Carlo procedure~\cite{lapeyre1998} 
routinely implemented in Particle-In-Cell (PIC) codes  to account for high-energy photon emission 
and its back-reaction on the electron dynamics~\cite{duclous2011,lobet2016}. 

In a recent work, Ref.~\cite{FNiel}, it was shown that for $\chi \lesssim 1$ (and arbitrary geometry of interaction), 
this discontinuous jump process is well approximated by a diffusion process.
Performing a Fokker-Planck expansion of the {\it collision operator} in the limit $\gamma_{\gamma} \ll \gamma$, 
it was shown that the electron momentum obeys a stochastic differential equation with 
a deterministic term containing both the Lorentz force and {\it quantum corrected} Landau-Lifshitz radiation friction force~\cite{Landau_CED} 
and a (stochastic) diffusion term accounting for the stochasticity of high-energy photon emission inherent to its quantum nature.
This work allowed us to bridge the quantum (MC) and deterministic [quantum-corrected Landau-Lifshitz (cLL)] descriptions of radiation reaction.
It also provided us with an additional description, henceforth referred to as the Fokker-Planck (FP) description, of radiation reaction 
to complement the standard MC and cLL descriptions. We then computed the equation of evolution of the successive (energy) moments of the electron distribution function
for all three descriptions. Doing so, we were able to show that the average energy was well reproduced by all three models. The FP or MC were needed to reproduce
correctly the energy spread (second order moments) while the third order moment was predicted correctly only by the MC method. 
Using this, we investigated the domain of validity of the various descriptions to correctly model the evolution of the electron distribution function.
In next Section, we will focus on the ability of the various models to correctly describe the spectral properties of the radiation associated to RR.

\subsection{High energy photon spectrum accounting for radiation reaction}\label{sec:rad_spectrum:C}

The Master Eqs~\eqref{eq:Master1} and~\eqref{eq:Master2} capture the physical picture of RR being the cumulative effect
of the successive and incoherent emissions of high-energy photons by the radiating electrons.
Hence, the instantaneous photon energy spectrum can be obtained at any given time $t$
by multiplying Eq.~\eqref{eq:Master2} by $\gamma_{\gamma}$, leading: 
\begin{eqnarray}\label{eq:dPrad_RR}
\frac{dP^{\rm RR}_{\rm{rad}}}{d\gamma_{\gamma}}(t,{\bf x},\gamma_{\gamma},{\bf \Omega}) =\!\! \int_1^{+\infty}\!\!\!\!d\gamma\, \frac{dP_{\rm inst}}{d\gamma_{\gamma}} \,f_e(t,{\bf x},\gamma,{\bf \Omega})\,,
\end{eqnarray}
where $dP_{\rm inst}/d\gamma_{\gamma}$ is the (single electron) instantaneous power spectrum given by Eq.~\eqref{eq:spectrum}.
The remarkable thing here is that, as the effect of RR is encompassed into the electron electron distribution function $f_e$, 
Eq.~\eqref{eq:dPrad_RR} returns the spectrum of high-energy photons accounting for the effect of RR.
It can be computed at any given time $t$ provided that the electron distribution function is known at this time.
The remarkable thing here is that the final radiated spectrum (with RR) depends only on the instantaneous electron distribution function. 
In other words, all the effects of the RR will materialise in $f_e$.
This allows us to readily apply all of the predictions on the electron distribution function discussed in Ref.~\cite{FNiel} to assess the effects of RR on the high-energy photon spectrum.
In the next Sec.~\ref{sec:results}, we will in particular demonstrate that both the MC and FP descriptions for the evolution of the electron populations, 
which were shown to correctly describe the evolution of the first two energy moments of the electron distribution function (namely the electron mean energy and energy dispersion)
provide similar predictions for the emitted photon spectra, while the cLL description fails in predicting the correct photon spectrum at large $\chi$.

Let us finally note that, interpreting $f_e$ not as a distribution function for a set (population) of electrons but 
as the probability density function for a single electron to have an energy between $\gamma$ and $\gamma + d\gamma$ at a time $t$,  
Eq.~\eqref{eq:dPrad_RR} can also be employed to compute the radiation spectrum of a single electron taking into account RR.

\section{Numerical simulations}\label{sec:results}
In what follows we first detail the numerical method developed in the PIC code {\sc Smilei} to simulate the evolution of a radiating electron population 
in a given external electromagnetic accounting for RR, focusing in particular on how we extract the emitted radiation properties.
We then present the results of simulation considering both the head-on collision of an ultra-relativistic electron beam with an electromagnetic plane-wave,
and the evolution of a hot electron population in a constant and homogeneous magnetic field.

\subsection{Method}

To investigate the properties of high-energy photons emitted by radiating ultra-relativistic electrons in various configurations, 
two series of simulations have been performed  with the PIC code {\sc Smilei}~\cite{smilei}. 
This open-source code has been upgraded to allow for the treatment of radiation reaction either by considering 
the quantum-corrected Landau-Lifshitz (cLL) force acting on an electron population, 
the Fokker-Planck (FP) pusher developed in Ref.~\cite{FNiel} which complement the deterministic cLL description by adding a stochastic diffusion term,
or the full Monte-Carlo (MC) procedure allowing to generate high-energy (macro-)photons.
The implementation of these three descriptions closely follows what is presented in Ref.~\cite{FNiel} (see also Refs.~\cite{duclous2011,lobet2016} for the cLL and MC descriptions).
It has been improved to allow for a better handling of the vectorization capabilities of new high-performance super-computers,
but these improvements are beyond the scope of this work and will be presented elsewhere.

From these simulations, we have been able to extract the spectral properties (energy and angular distributions) of the emitted radiation.
This is particularly simple when using the MC procedure as (macro-)photons are naturally created at runtime.
The spectral properties of the radiated light can then be easily reconstructed (at each time-steps or integrated over the full simulation time) 
by {\it depositing} the (macro-)photons energy over a regular grid in energy (and/or angle).
When considering the two other (cLL or FP) descriptions, the spectral properties of the emitted radiation where computed, at runtime, considering
that all electrons emit, at each timestep and in their direction of propagation ${\bf\Omega}$, the full spectrum given by Eq.~\eqref{eq:spectrum}.
Let us recall here that the descriptions of radiation reaction used in this work rely on the LCFA.

This procedure allowed us to access the spectral properties of the emitted radiation, accounting for radiation reaction, in various configurations. 
In what follows, we detail the results obtained considering 
(i) an ultra-relativistic electron bunch head-on collision with an electromagnetic plane-wave (Sec.~\ref{sec:numResultsPW}) and
(ii) an a hot electron population (broad zero-drift Maxwell-J\"uttner distribution) radiating in a constant magnetic field (Sec.~\ref{sec:numResultsMJ}).
In all these simulations, the electron feedback on the external electromagnetic field was turned off (no-current deposition) to focus on the particle dynamics
in the external field without accounting for Coulomb repulsion.
The timestep and number of macro-particles were chosen to ensure convergence of the simulations and will be detailed for each case.

\subsection{Results}

In what follows, we will consider two initial set-up for the electron populations.
First (Sec.~\ref{sec:numResultsPW}), an ultra-relativistic electron beam similar to that investigated in Ref.~\cite{FNiel} is considered,
following a Maxwell-J\"uttner distribution with drift velocity ${\bf v}_d=c\,\sqrt{\gamma_0^2-1}/\gamma_0 \hat{\bf x}$ 
with $\gamma_0 = 1800$ and (proper) temperature $T_0 = 2.5 \times 10^{-3}~m c^2$.
Second (Sec.~\ref{sec:numResultsMJ}), a hot electron population with zero-drift Maxwell distribution and temperature $T_0 = 600~m c^2$ [corresponding to
an initial average electron Lorentz factor $\gamma_0 = \langle\gamma\rangle(t=0) = 1800$] is considered.
Note that, in {\sc Smilei}, the correct loading of the relativistic Maxwell-J\"uttner distribution is ensured following the method proposed in Ref.~\cite{zenitani2015}.

\subsubsection{Electron beam head-on collision with an UHI plane-wave}\label{sec:numResultsPW}

 \begin{figure}
\begin{center}
\includegraphics[width=\textwidth]{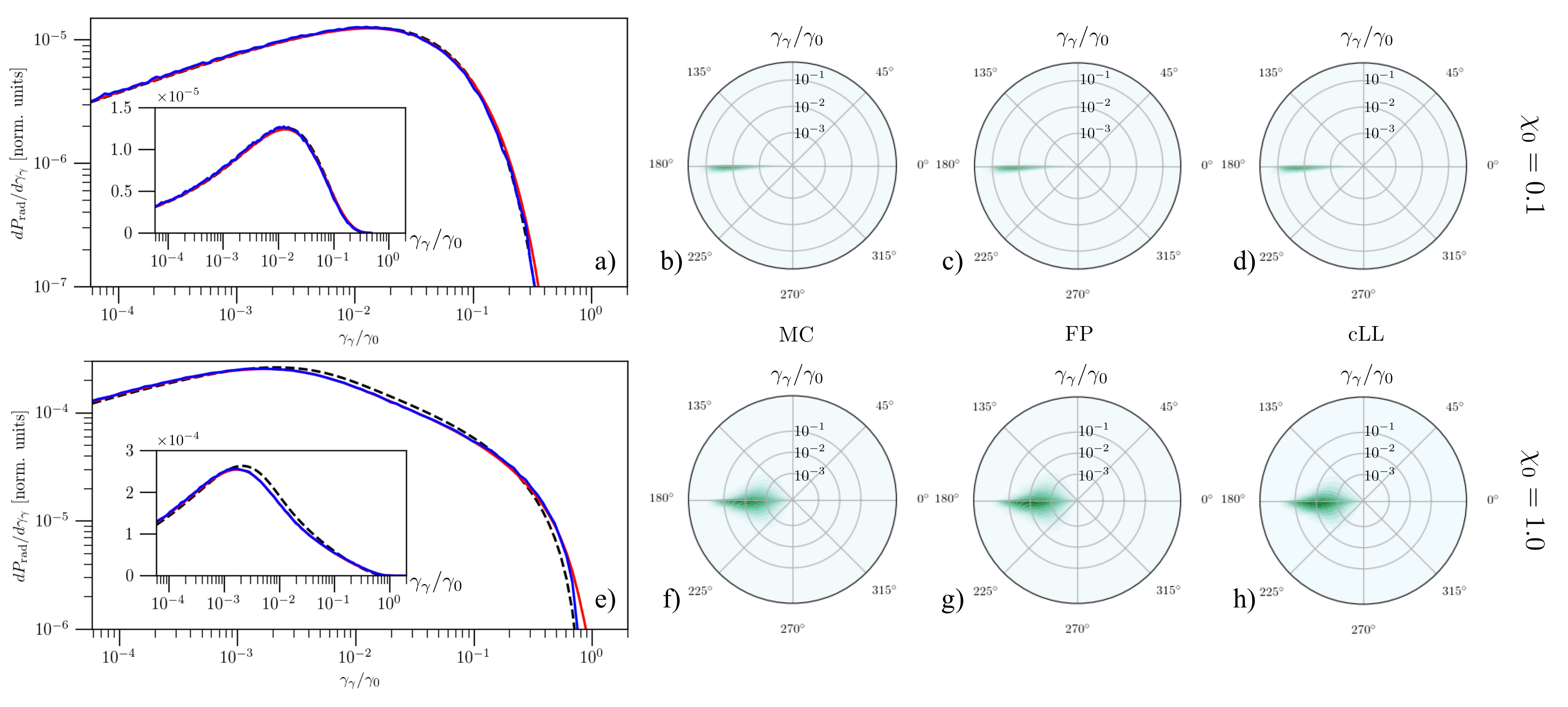}
\caption{Time integrated high-energy photon spectra considering a high-energy electron bunch head-on collision with a large amplitude electromagnetic plane wave. Top panels (a-d) correspond to an initial electron quantum parameter $\chi_0 = 0.1$, bottom panels (e-h) to $\chi=1.0$. Black dashed lines correspond to the deterministic, quantum-corrected Landau-Lifshitz (cLL) simulations, red lines to the stochastic Fokker-Planck (FP) simulations and blue lines to the Monte-Carlo (MC) simulations.  Left panels (a,e) report the energy spectra integrated over the full simulation time (the $y$-axis is reported in linear scale in the inserts). Panels (b,f) show the high-energy photon energy-angle-distributions computed using the MC method. Panels (c,g) show the high-energy photon energy-angle-distributions computed using the FP method. Panels (d,h) show the high-energy photon energy-angle-distributions computed using the cLL method.  }\label{fig:PWint} 
\end{center}
\end{figure}

 \begin{figure}
\begin{center}
\includegraphics[width=\textwidth]{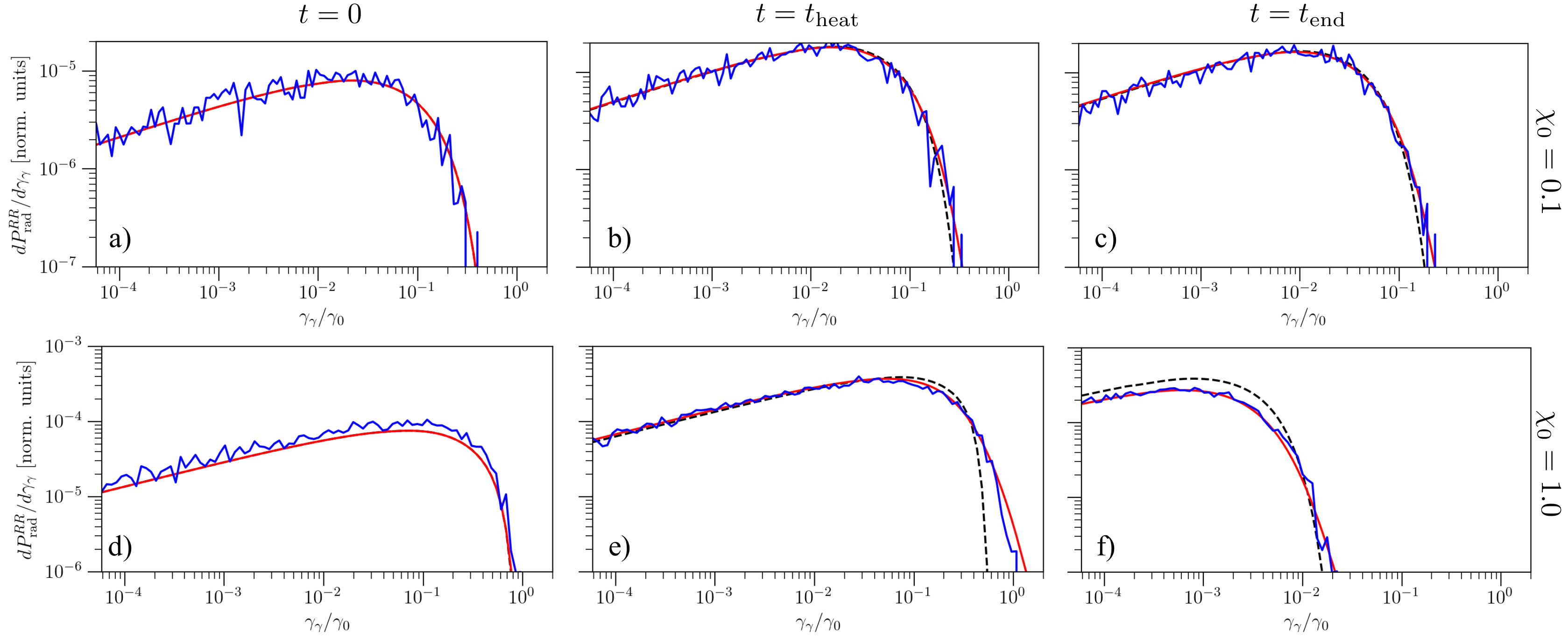}
\caption{Instantaneous high-energy photon spectra considering a high-energy electron bunch head-on collision with a large amplitude electromagnetic plane wave. Top panels (a-c) correspond to an initial electron quantum parameter $\chi_0 = 0.1$, bottom panels (d-f) to $\chi=1.0$.  Color-coding is the same than for Fig.~\ref{fig:PWint}. Spectra are reported at the early time of interaction $t=0$ (a,d), at the time of maximum energy spread of the electron distribution function [$t_{\rm heat} \simeq 100\,\omega_0$ for $\chi_0=0.1$ (panel b) and $t_{\rm heat} \simeq 4.4\,\omega_0$ for $\chi_0=1.0$ (panel e)], and at the end of the simulation (c,f). }\label{fig:PW} 
\end{center}
\end{figure}

We start by simulating the interaction of an ultra-relativistic electron beam, following a Maxwell-J\"uttner distribution with temperature $T_0 = 2.5 \times 10^{-3}~m c^2$ 
and drift velocity ${\bf v}_d$ corresponding to the Lorentz factor $\gamma_0 = 1800$, colliding head-on with a large amplitude, linearly polarized electromagnetic plane-wave (henceforth 
referred to as the laser). 
The amplitude of this plane-wave is chosen so that $\chi_0 = 2 a_0 \gamma_0\,\tau_e\omega_0/\alpha= 10^{-1}$ and $1$ (corresponding to normalized vector potential $a_0 = 11.4$ and $114$, respectively,
and considering $\lambda_0 = 1~{\rm \mu m}$ the laser wavelength). 
The end of the simulation is taken when the energy decrease of the electron population has considerably slowed down (i.e. we approach a regime in which radiation losses are not important).
We used $t_{\rm end} \simeq 190/\omega_0$ for $\chi_0 = 0.1$ and $t_{\rm end} \simeq 60/\omega_0$ for $\chi_0 = 1$, with $\omega_0=2\pi c/\lambda_0$ the laser angular frequency. 

For both simulations, the electron distribution function evolves exactly as reported in Fig.~12 of Ref.~\cite{FNiel}.
The simulation results concerning the radiation spectra are summarized in Figs.~\ref{fig:PWint} and Fig.~\ref{fig:PW} considering the time-integrated or instantenous spectra, respectively.
Let us start by noticing that all three descriptions (cLL, FP and MC) predict similar the time-integrated energy and angular distributions (Fig.~\ref{fig:PWint}).
Note also that, as expected for this interaction configuration, the photon emission is beamed in the direction of the electron beam ($\theta \sim -180^{\circ}$,
with $\theta$ the angle of emission in the laser polarization plane). 
In particular, for the case $\chi_0 = 0.1$, all three descriptions provide the very same prediction for the time-integrated radiation energy spectrum (Fig.~\ref{fig:PWint}a). 
Discrepancies can be seen when $\chi_0 = 1$ between the FP/MC one the one hand, and the (deterministic) cLL description on the other hand.
In particular, we note a hardening of the photon spectrum (increase of the radiated power at high energies) for the FP/MC predictions, while the (deterministic) cLL one
overestimates the power spectrum at intermediate photon energies.
Both stochastic descriptions (FP and MC) are however in very good agreement for both initial values of $\chi_0$. 
A small discrepancy for the highest energy photons can still be observed in between these two methods which follows from the spurious up-scattering of electrons appearing at high-$\chi$ in the FP model (see Ref.~\cite{FNiel}). Still, the discrepancies in between the two stochastic methods are much more tenuous than those observed in the electron distribution functions reported in Ref.~\cite{FNiel}.

In order to study more precisely the radiation spectrum, we now turn to the instantaneous spectra reported at different times in Fig.~\ref{fig:PW}.
At $t = 0$,  Fig.~\ref{fig:PWint} (panels a and d), the three models give the exact same spectrum. 
This is due to the fact that the total spectrum only depends on the electron distribution function [see Eq.~\eqref{eq:dPrad_RR}], which is the same initially in the three simulations.
At later times, Fig.~\ref{fig:PWint} (panels b and e), differences can be clearly observed between the FP/MC predictions and those of the cLL description for $\chi_0 =1$ (panel e), while
all three method lead to the same instantaneous spectra at lower $\chi_0=0.1$ (panel b).
The particular time $t=t_{\rm heat}$ reported in panels b and e correspond to the time at which the energy dispersion of the electron population (second order moment of
the electron distribution function) is maximal (see our discussion in  Ref.~\cite{FNiel} as well as Refs.~\cite{vranic2016,ridgers2017}). 
This spread of the electron distribution follows from the stochastic nature of high energy photon emission (see also Ref.~\cite{Neitz}). 
While it is correctly described by both the FP and MC approaches, it is not accounted for in the deterministic (cLL) approach which only predict a narrowing (cooling) of the electron distribution function~\cite{tamburini2010,FNiel}.
Due to this increased energy spread (quantum stochastic heating), we have at $t=t_{\rm heat}$ a non-negligible number of electron with
$\gamma > \langle \gamma \rangle$ in the FP/MC models with respect to the deterministic cLL model allowing for the emission of very high energy photons (hence the observed hardening of the radiation). We note for $\chi_0=1$ a slight overestimate with the FP method of the power emitted at large photon energy, which as discussed above, follows from the FP-inherent spurious electron up-scattering.
In addition, the number of particles with an energy $\gamma \sim \langle \gamma \rangle$ is larger in the cLL model than in the FP/MC models and we observe a spectrum which is more intense
at intermediate energy in the cLL descriptions than in the FP/MC ones. 
Similarly, we find more electrons at lower energy $\gamma < \langle\gamma\rangle$ in the FP/MC models than in the cLL model, and the energy yield at these low energies is larger considering the 
stochastic models (FP \& MC) than considering the deterministic one (cLL). 
Finally, at $t = t_{\rm end}$, we make the same observations as for $t = t_{\rm heat}$, with the additional particularity that the MC and FP models are now in perfect 
agreement, which is coherent with the analysis carried in Ref.~\cite{FNiel} that the FP and MC models, while yielding different predictions at high $\chi$ and small time
 are in good agreement at latter times.
 
To conclude with this first series of simulations, we would like to point out that the discrepancies in the emitted radiation spectra predicted by all three methods are much more tenuous than those 
observed in the electron distributions as discussed in Ref.~\cite{FNiel}, in particular when considering time-integrated spectra. 
Furthermore, while some large discrepancies on the highest (>2) energy moments of the electron distribution were reported in between the FP and MC predictions for large $\chi_0 \simeq 1$, 
both FP and MC methods are here demonstrated to provide interestingly similar predictions on the emitted radiation properties. This indicates that the emitted radiation properties depend mainly on
 the first two energy moments of the electron distribution function.

\subsubsection{Hot (Maxwell-J\"uttner) electron population radiating in a constant magnetic field}\label{sec:numResultsMJ}

We now simulate the evolution of a hot Maxwell-J\"uttner electron population [with zero drift and temperature $\theta = T/(mc^2) = 600$]
in a constant-uniform magnetic field with magnitude corresponding to initial average electron quantum parameters 
$\chi_0 = \langle\chi\rangle(t=0) = 10^{-1}$ and $1$ (correspondingly, $225~\rm{kT}$ and $2.25~\rm{MT}$). 

\begin{figure}
\begin{center}
\includegraphics[width=\textwidth]{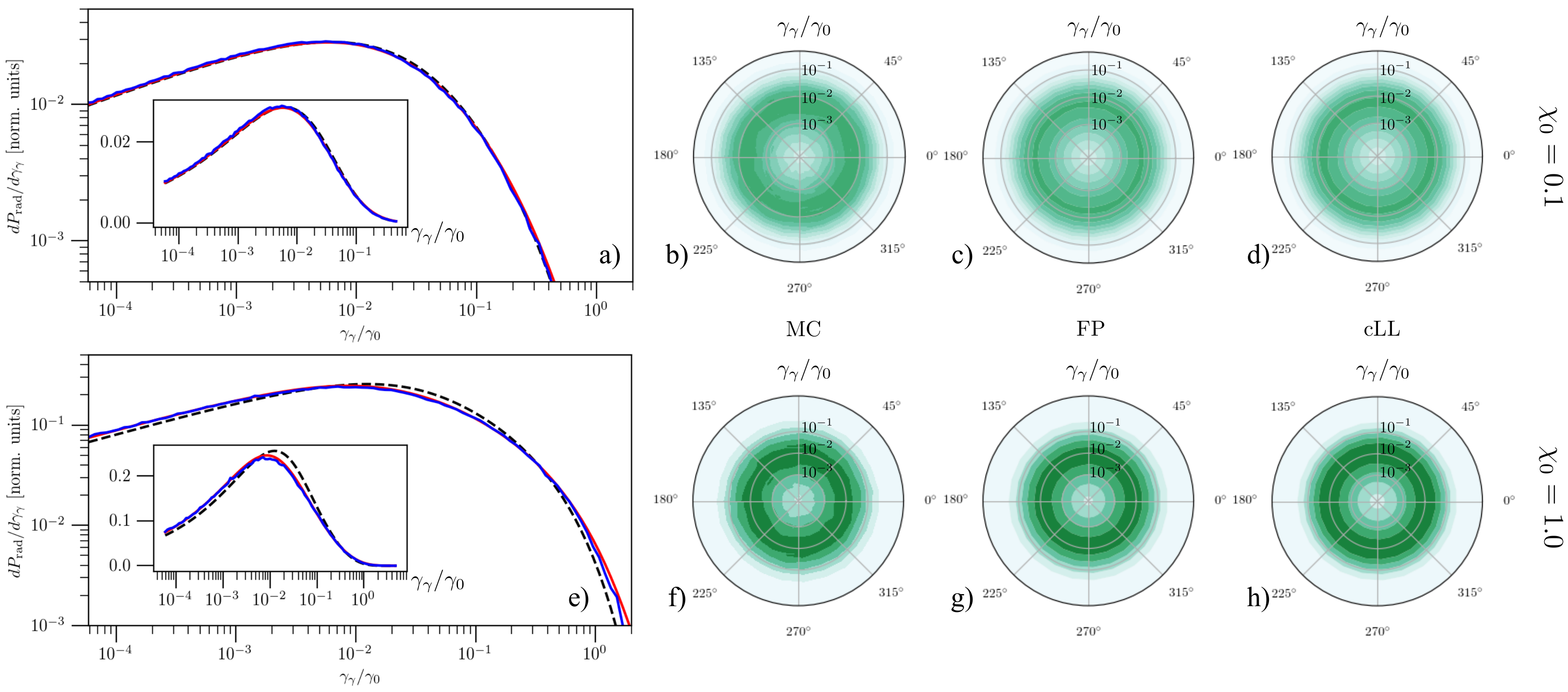}
\caption{Time integrated high-energy photon spectra considering a hot (Maxwell-J\"uttner) electron population radiating in a constant magnetic field. Top panels (a-d) correspond to an initial electron quantum parameter $\chi_0 = 0.1$, bottom panels (e-h) to $\chi=1.0$. Color-coding is the same than for Fig.~\ref{fig:PWint}. Left panels (a,e) report the energy spectra integrated over the full simulation time (the $y$-axis is reported in linear scale in the inserts). Panels (b,f) show the high-energy photon energy-angle-distributions computed using the MC method. Panels (c,g) show the high-energy photon energy-angle-distributions computed using the FP method. Panels (d,h) show the high-energy photon energy-angle-distributions computed using the cLL method.  }\label{fig:MJint} 
\end{center}
\end{figure}

 \begin{figure}
\begin{center}
\includegraphics[width=\textwidth]{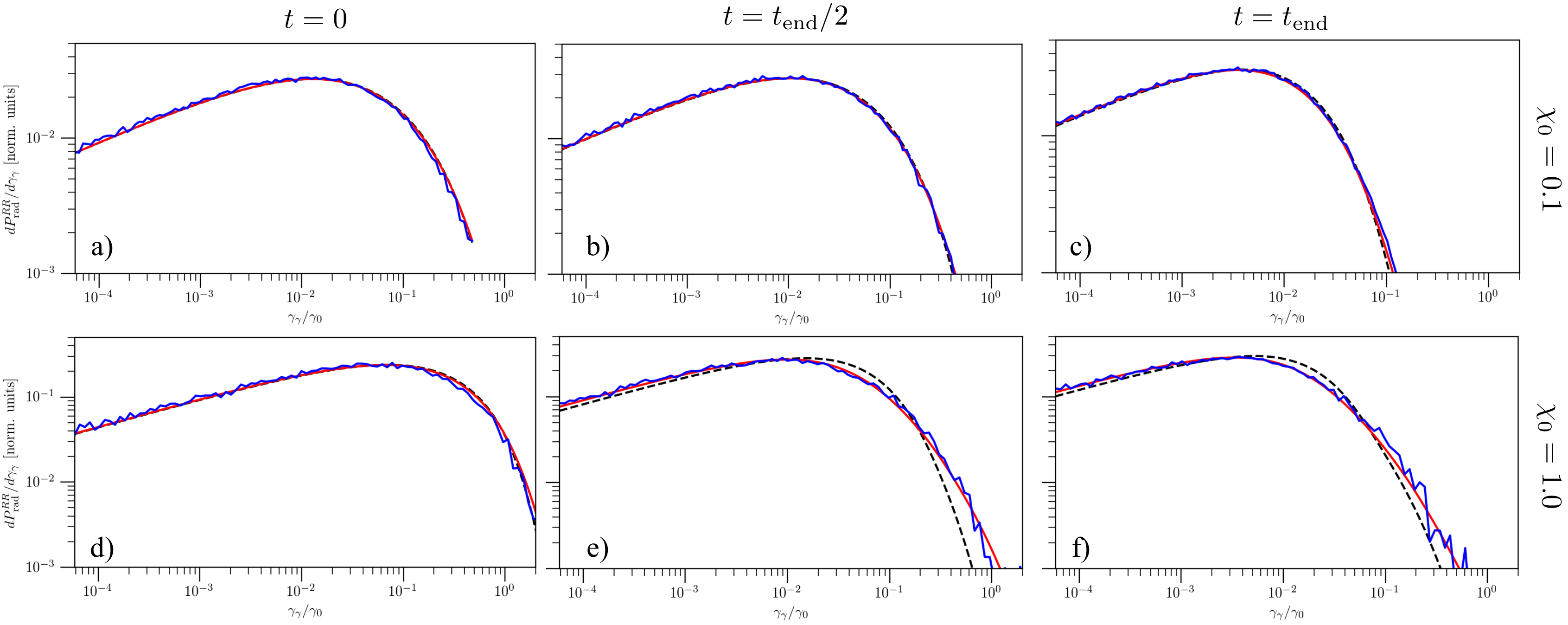}
\caption{Instantaneous high-energy photon spectra considering a hot (Maxwell-J\"uttner) electron population radiating in a constant magnetic field. Top panels (a-c) correspond to an initial electron quantum parameter $\chi_0 = 0.1$, bottom panels (d-f) to $\chi=1.0$. Color-coding is the same than for Fig.~\ref{fig:PWint}.  Spectra are reported at the early time of interaction $t=0$ (a,d), at half of the simulation (b,e), and at the end of the simulation (c,f).}\label{fig:MJ} 
\end{center}
\end{figure}

The duration of the simulations was here taken arbitrarily as $t_{\rm end} \simeq 235/\omega_c$ for $\chi_0=0.1$, and $t_{\rm end} \simeq 16/\omega_c$ for $\chi_0=1$,
 with $\omega_c = eB/(m \gamma_0)$ the synchrotron frequency for an electron with the average Lorentz factor $\gamma_0 = \langle\gamma\rangle(t=0)=1800$.
As shown in Ref.~\cite{FNiel}, for this particular case of a broad Maxwell-J\"uttner distribution,
 all three descriptions predict a cooling (narrowing) of the electron distribution function and the reported simulations were long enough for the electron populations to significantly cool down. 
 The simulation results are presented in Figs.~\ref{fig:MJint} and~\ref{fig:MJ}.
For this particular simulations, and similarly to what was reported in the previous Section, all three models are in perfect agreement for $\chi_0 = 0.1$ and only slight deviations are observed at $\chi_0=1$ in the time-integrated spectra (Figs.~\ref{fig:MJint}, note however the isotropic the angular distribution in the plane transverse to the magnetic fields which follows from the symmetry of the system, ${\bf v}_d=0$). The excellent agreement at all times is confirmed in Fig.~\ref{fig:MJ} for $\chi_0=0.1$
while discrepancies at larger initial quantum parameter $\chi_0=1$ are found to follow the same line as observed in the previous Sec.~\ref{sec:numResultsPW}.
Again, in light of our previous study on the successive moments of the electron distribution function, and in particular in the ability of only the FP and MC descriptions to correctly capture the
temporal evolution of the electron population first two moments, the excellent agreement between FP and MC simulations observed here once more indicate that the emitted radiation spectrum 
is mainly sensitive to this first two moments.

\section{Conclusions}\label{sec:conclusion}

In this paper we showed that the radiated spectrum in laser-electron beam interaction or in the evolution of a hot plasma in a constant uniform magnetic field accounting for RR effects
can be easily reconstructed, in the LCFA, from the instantaneous electron distribution function. We can then apply predictions on the electron distribution function~\cite{FNiel} on the radiated spectrum. In particular, a reduced description of the distribution function evolution involving two energy moments -when the Fokker Planck modeling is appropriate- or three energy moments 
-when the Monte-Carlo modelling is necessary- of $f_e$ 
can be used in principle. In practice, the excellent 
agreement of the FP model with the MC procedure up
to $\chi_0 = 1$ shows
that only the first two moments are actually sufficient 
to describe the spectrum the qualitative features of 
the spectrum.
This is verified by PIC simulation. A remarkable result is that, even when the quantum parameter is close to one, the radiated spectrum smooths out differences in the models necessary to correctly describe the electron distribution function. As a consequence, the range of parameters for wich the prediction on the spectrum given by the Fokker-Planck description is correct becomes very wide. In order to discriminate experimentally between different models and evidence more easily pure quantum effects direct observation of the electron distribution function is advantageous.

\section*{Acknowledgements}
The authors thank the {\sc Smilei} development team for technical support.
Financial support from PALM (ANR-10-LABX-0039-PALM, Junior Chair SimPLE) and Plas@Par (ANR-11-IDEX-0004-02-Plas@Par) is acknowledged.
This work was granted access to the HPC resources from GENCI-TGCC (Grant 2017-x2016057678).

\section*{References}

\end{document}